\documentclass[USenglish,twocolumn]{article}

\ifx\directlua\undefined\ifx\XeTeXcharclass\undefined
  \usepackage[utf8]{inputenc}                           
  \else\RequirePackage[no-math]{fontspec}[2017/03/31]\fi 
  \else\RequirePackage[no-math]{fontspec}[2017/03/31]\fi 
\usepackage[sort&compress,square,numbers]{natbib}
\usepackage[big,online]{dgruyter}
\usepackage{subfig}

\theoremstyle{dgthm}

\theoremstyle{dgdef}

\begin{document}

\articletype{Research Article}

\author[2]{Kai Schwennicke}
\author[3]{Noel C. Giebink}
\author*[1]{Joel Yuen-Zhou} 

\affil[1]{Department of Chemistry and Biochemistry, University of California San Diego, La Jolla, CA 92093, USA, E-mail: joelyuen@ucsd.edu. https://orcid.org/0000-0002-8701-8793}
\affil[2]{Department of Chemistry and Biochemistry, University of California San Diego, La Jolla, CA 92093, USA, E-mail: kschwenn@ucsd.edu.}
\affil[3]{Department of Electrical Engineering and Computer Science, and Physics, University of Michigan, Ann Arbor, Michigan, 48109, USA, E-mail: ngiebink@umich.edu. https://orcid.org/0000-0002-3798-5830}

\title{Extracting accurate light-matter couplings from disordered polaritons}
\runningtitle{Extracting accurate light-matter couplings from disordered polaritons}
\abstract{The vacuum Rabi splitting (VRS) in molecular polaritons stands as
a fundamental measure of collective light-matter coupling. Despite
its significance, the impact of molecular disorder on VRS is not fully
understood yet. This study delves into the complexities of VRS amidst
various distributions and degrees of disorder. Our analysis provides
precise analytical expressions for linear absorption, transmission,
and reflection spectra, along with a ``sum'' rule, offering a straightforward
protocol for extracting accurate collective light-matter coupling
values from experimental data. Importantly, our study cautions against
directly translating large VRS to the onset of ultrastrong coupling
regime. Furthermore, for rectangular disorder, we witness the emergence
of narrow side bands alongside a broad central peak, indicating an
extended coherence lifetime even in the presence of substantial disorder.
These findings not only enhance our understanding of VRS in disordered
molecular systems but also open avenues for achieving prolonged coherence
lifetimes between the cavity and molecules via the interplay of collective
coupling and disorder.}

\journalyear{2024}

\maketitle

\section{Introduction} 

In the realm of molecular polaritons, the phenomenon of vacuum Rabi
splitting (VRS) stands as an established metric for gauging the strength
of collective light-matter coupling. Traditionally, this interaction
is classified into several regimes: weak, strong, ultrastrong, and
deep-strong \cite{frisk2019ultrastrong}. In particular, molecular
polaritons are often observed in the realm of strong coupling, with
a wide range of potential applications such as catalysis \cite{hutchison2012modifying,thomas2016ground,thomas2019tilting,ahn2023modification},
exciton transport \cite{rozenman2018long,pandya2022tuning,engelhardt2022unusual,engelhardt2023polariton,xu2023ultrafast,balasubrahmaniyam2023enhanced},
and Bose-Einstein condensation \cite{kasprzak2006bose,daskalakis2014nonlinear,plumhof2014room}.
While reaching the ultrastrong coupling regime remains experimentally
challenging, there are a number of experiments that have pushed the
limits of molecular systems into this intriguing regime \cite{schwartz2011reversible,mazzeo2014ultrastrong,george2016multiple,eizner2018organic}.

In the idealized scenario of $N$ identical molecules strongly coupled
to a single photonic mode, the magnitude of VRS scales linearly with
$\sqrt{N}$ \cite{tavis1968exact,ribeiro2018polariton}. However,
the inherent complexity of molecular ensembles introduces a compelling
challenge, as molecular disorder becomes an inescapable feature. Molecular
disorder exerts a profound influence on various aspects of polariton
physics, including transport \cite{feist2015extraordinary,schachenmayer2015cavity,botzung2020dark,allard2022disorder,pandya2022tuning,xu2023ultrafast,engelhardt2022unusual,suyabatmaz2023vibrational},
photoconductivity \cite{krainova2020polaron}, photoreactivity \cite{perez2023frequency},
and vibropolaritonic chemistry \cite{du2022catalysis}. Surprisingly, even though the effects of disorder where theoretically studied early on \cite{pau1995microcavity,savona1995quantum,houdre1996vacuum,pau1996theory},
the effects of disorder on VRS splitting are still a debate within
the community. Early explorations by Houdr{\'e} et al. \cite{houdre1996vacuum}
suggested that disorder (or inhomogeneous broadening) should have
no impact on the size of the splitting. However, these conclusions
have been contested in recent investigations \cite{sommer2021molecular,engelhardt2022unusual,gera2022effects,gera2022exact,cohn2022vibrational,zhou2023interplay,climent2023kubo}
which note that disorder can both enhance and suppress the VRS.
\begin{figure*}[!ht]
\subfloat[\label{fig:linear_spec}]{\includegraphics[width=1\columnwidth]{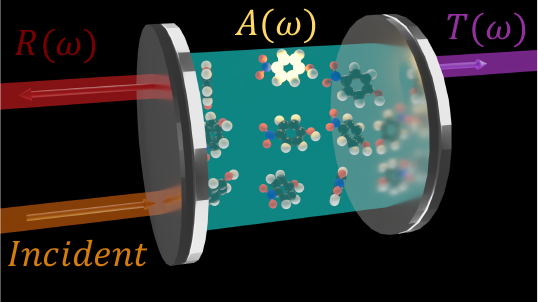}

}\subfloat[\label{fig:level_repulsion}]{\includegraphics[width=1\columnwidth]{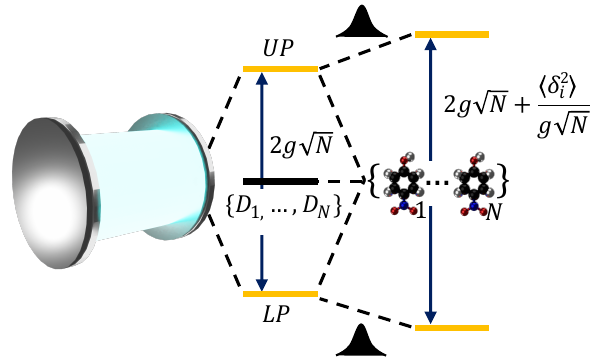}

}\caption{(A) Linear spectroscopy of molecular polaritons as absorption $A(\omega)$,
transmission $T(\omega)$, and reflection $R(\omega)$ (B) For weak
disorder, a perturbative approach to understand the role of disorder
is useful. The zeroth-order Hamiltonian can be taken to be the disorderless
system, comprising of the photon mode interacting with the $N$
degenerate molecules to form the upper (\emph{UP}) and lower (\emph{LP})
polaritons, alongside $N-1$ dark states ($\{D_{1},...,D_{N-1}\}$).
The VRS at resonance in this case happens to be $2g\sqrt{N}$, where
$g$ is the single molecule light-matter coupling. The molecular disorder
then perturbatively couples the polaritons to the manifold of dark
states, inducing level repulsion between the polaritons, as discussed
in Refs. \cite{sommer2021molecular,cohn2022vibrational}, and \cite{zhou2023interplay},
thus increasing the VRS. This repulsion depends on the variance of
the disorder distribution ($\langle\delta_{i}^{2}\rangle$, where
$\delta_{i}=\omega_{i}-\omega_{0}$ and $\omega_{0}$ represents the
center of the distribution). Given the perturbative character of this
analysis, it does not apply to strong disorder.}
\end{figure*}

In this article, we revisit the problem of VRS and disorder, embarking
on a comprehensive study of linear absorption $A$, transmission $T$,
and reflection $R$ properties of molecular polaritons, considering
various distributions and magnitudes of disorder. Our aim is not only
to elucidate the intricate interplay between molecular disorder and
VRS, but also to provide a robust method for accurately extracting
light-matter coupling parameters. Our findings align with recent reports,
demonstrating that VRS tends to increase with disorder, reaches a
saturation point, and eventually decreases to zero for a wide range
of disorder distributions. Note that $A,T,R$ do not in general give
the same value of VRS \cite{savona1995quantum}. Hence, while some
of the results are already known in the literature, we deem it valuable
to collect all the results in a single study. Significantly for experiments,
we unveil what seems to be a ubiquitous scenario: the presence of
substantial disorder can dramatically enhance the VRS leading to an
\emph{apparent} onset of the ultrastrong coupling regime, despite
the underlying collective light-matter coupling firmly residing within
the strong coupling regime. Moreover, we introduce a novel sum rule
that proves instrumental in extracting precise values of collective
light-matter coupling, particularly when both absorption and transmission
can be measured in the experimental setup. Crucially, this sum rule
demonstrates generality across all types of disorder. In the case
of a rectangular distribution (which was briefly discussed in Ref.
\cite{gera2022effects}), we observe the emergence of two narrow polariton
peaks in the spectra, reminiscent of the pronounced spectral narrowing
witnessed in the context of surface lattice resonances \cite{kravets2018plasmonic}
and other phenomena associated with Wood anomalies \cite{wood1902xlii,mcmahon2007tailoring,de2007colloquium,darweesh2018role,galiffi2020wood}.

\section{Results}

\subsection{Model}
\begin{table*}
\centering
\caption{Real and imaginary parts of the molecular susceptibility\label{tab:Real-and-imaginary}}
\begin{tabular}{cccc}
 & \multicolumn{3}{c}{Susceptibility}\\
\cline{2-4} 
\noalign{\vskip0.1cm}
Real or Imaginary & Gaussian\textsuperscript{a} & Lorentzian & Rectangle \\
\midrule 
\noalign{\vskip0.5cm}
$\chi'(\omega)$ & $-\frac{g^{2}N\sqrt{2}}{\sigma}F(\frac{\omega-\omega_{0}}{\sqrt{2}\sigma})$ & $-g^{2}N\frac{(\omega-\omega_{0})}{(\omega-\omega_{0})^{2}+(\sigma/2)^{2}}$ & $-\frac{g^{2}N}{\sigma}\ln|\frac{\omega-\omega_{0}+\sigma/2}{\omega-\omega_{0}-\sigma/2}|$ \\
\noalign{\vskip0.5cm}
$\chi''(\omega)$ & $\frac{g^{2}N}{\sigma}\sqrt{\frac{\pi}{2}}\exp[-\frac{1}{2}(\frac{\omega-\omega_{0}}{\sigma})^{2}],$ & $g^{2}N\frac{\sigma/2}{(\omega-\omega_{0})^{2}+(\sigma/2)^{2}}$ & $\frac{g^{2}N\pi}{\sigma}\text{rec}[2(\omega-\omega_{0})/\sigma],$ \\
\noalign{\vskip0.5cm}
\midrule
\noalign{\vskip0.1cm}
\multicolumn{4}{c}{\textsuperscript{a}Here $F(y)=\exp[-y^{2}]\int_{0}^{y}dte^{t^{2}}$ is
the Dawson function.}\\
\end{tabular}
\end{table*}

For concreteness, we consider $N$ two level systems coupled to a
single photon mode {[}in the rotating wave approximation (RWA), Tavis--Cummings
model \cite{tavis1968exact}{]}:
\begin{equation}
H=\hbar\omega_{ph}a^{\dagger}a+\sum_{i=1}^{N}\hbar\omega_{ex,i}\sigma_{i}^{\dagger}\sigma_{i}-\hbar\lambda\mu\Big(\sum_{i=1}^{N}a\sigma_{i}^{\dagger}+\text{h.c.}\Big),\label{eq:Hamiltonian}
\end{equation}
where $\omega_{ph}$ and $a$ are the photon frequency and annihilation
operator, $\omega_{ex,i}$ and $\sigma_{i}=|g_{i}\rangle\langle e_{i}|$
are the frequency and annihilation operator for the $i-$th two level
system, $\mu_{i}$ is the amplitude of the $i-$th transition dipole,
and $\hbar\lambda=\sqrt{\frac{\hbar\omega_{ph}}{2\epsilon_{0}\mathcal{V}}}$
is the vacuum field amplitude where $\epsilon_{0}$ is the vacuum
permittivity and $\mathcal{V}$ is the mode volume. In the thermodynamic
limit ($N\to\infty$), the linear absorption, transmission, and reflection
spectra are given by (see Refs. \cite{cwik2016excitonic} and \cite{yuen2023linear})
\begin{align}
A(\omega) & =\frac{2\kappa_{L}\chi''(\omega)}{|\omega-\omega_{ph}+i\frac{\kappa}{2}+\chi'(\omega)+i\chi''(\omega)|^{2}},\label{eq:polariton_absorption}\\
T(\omega) & =\frac{\kappa_{L}\kappa_{R}}{|\omega-\omega_{ph}+i\frac{\kappa}{2}+\chi'(\omega)+i\chi''(\omega)|^{2}},\label{eq:transmission}\\
R(\omega) & =1-A(\omega)-T(\omega).\label{eq:refleciton}
\end{align}
Here $\kappa=\kappa_{L}+\kappa_{R}$ is the total cavity decay rate,
and the respective decay rates into the left and right photon continua
are denoted by $\kappa_{L}$ and $\kappa_{R}$. The linear molecular
susceptibility $\chi(\omega)$ is given by 
\begin{equation}
\chi(\omega)=-\lim_{\gamma\to0^{+}}\sum_{i}^{N}\tanh\Big(\frac{\hbar\omega_{ex,i}}{2k_{B}T}\Big)\frac{|\lambda\mu_{i}|^{2}}{\omega-\omega_{ex,i}+i\frac{\gamma}{2}}.\label{eq:molecular_suceptabilty}
\end{equation}
Considering the case when $\hbar\omega_{ex,i}\gg k_{B}T$ and assuming
that all $N$ two level systems have the same transition-dipole amplitude
$\mu$, the molecular susceptibility becomes 
\begin{align}
\chi(\omega) & =-\lim_{\gamma\to0^{+}}g^{2}N\int d\omega_{ex}\frac{p(\omega_{ex})}{\omega-\omega_{ex}+i\frac{\gamma}{2}}\nonumber \\
&=\chi'(\omega)+i\chi''(\omega),\label{eq:molecular_suceptabilty_low_T}
\end{align}
where $g^{2}=|\lambda\mu|^{2}$ is the square of the single molecule light-matter coupling, $p(\omega_{ex})$ is the probability
distribution of excitation frequencies, and the real and imaginary
parts of the molecular susceptibility are $\chi'(\omega)=-g^{2}N\mathcal{P}\int d\omega_{ex}\frac{p(\omega_{ex})}{\omega-\omega_{ex}}$,
where $\mathcal{P}$ is the Cauchy principal value, and $\chi''(\omega)=g^{2}N\pi p(\omega)$.

To explore the effects of $p(\omega_{ex})$, we consider Lorentzian,
\begin{equation}
p(\omega_{ex})=\frac{1}{\pi}\frac{\sigma/2}{(\omega_{ex}-\omega_{0})^{2}+(\sigma/2)^{2}},\label{eq:Lorentzian}
\end{equation}
Gaussian, 
\begin{equation}
p(\omega_{ex})=\frac{1}{\sqrt{2\pi}\sigma}e^{-\frac{1}{2}(\frac{\omega_{ex}-\omega_{0}}{\sigma})^{2}},\label{eq:Gaussian}
\end{equation}
and rectangular
\begin{align}
p(\omega) & =\frac{1}{\sigma}\text{rec}[2(\omega-\omega_{0})/\sigma],\label{eq:rectangular_disorder}\\
\text{rec}[y] & =\begin{cases}
1, & |y|\leq1\\
0, & |y|>1
\end{cases}\label{eq:rectangular_function}
\end{align}
disorder. Table \ref{tab:Real-and-imaginary} lists the analytical
expressions of $\chi'(\omega)$ and $\chi''(\omega)$ for the three
different distributions.

\subsection{Lorentzian disorder}
\begin{figure*}
\subfloat[]{\includegraphics[width=0.666\columnwidth]{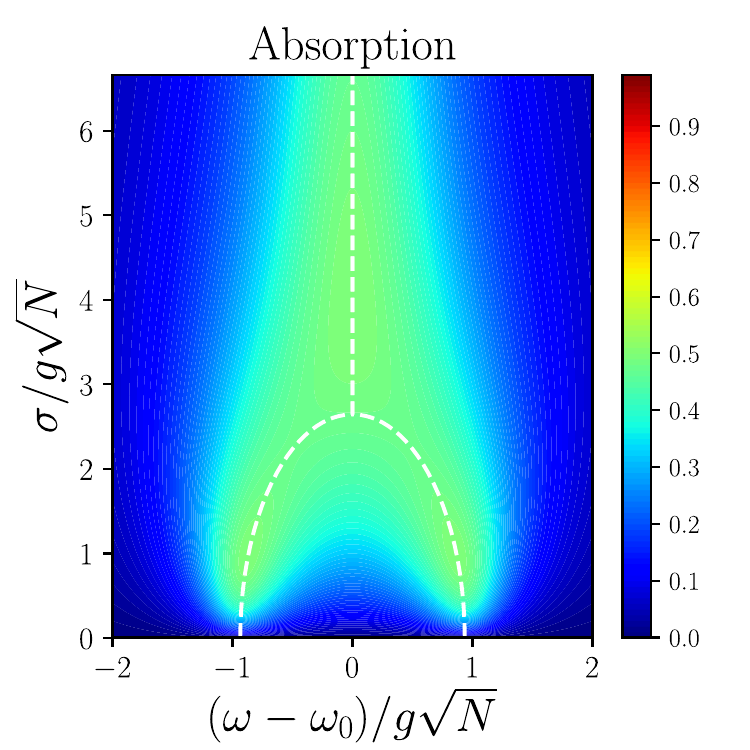}

}\subfloat[]{\includegraphics[width=0.666\columnwidth]{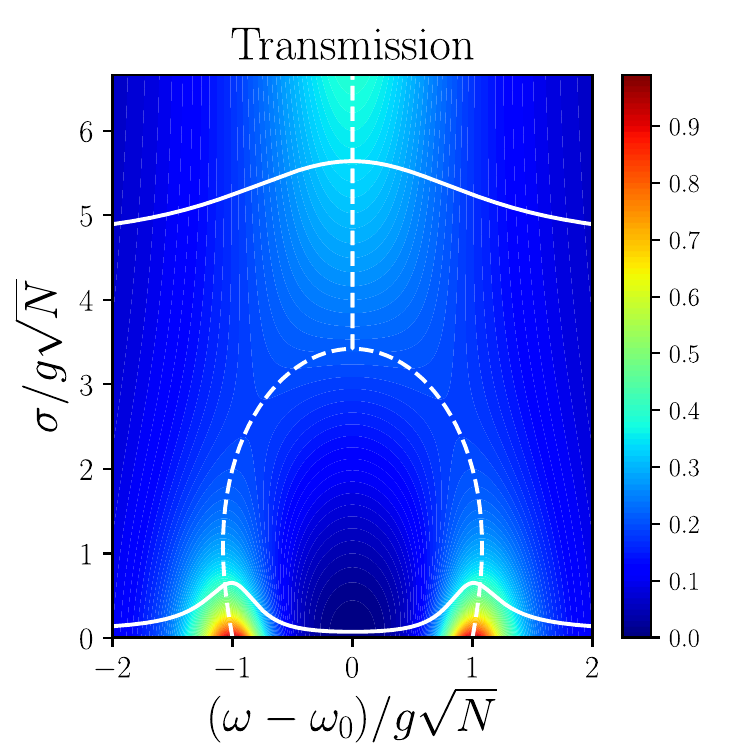}

}\subfloat[]{\includegraphics[width=0.666\columnwidth]{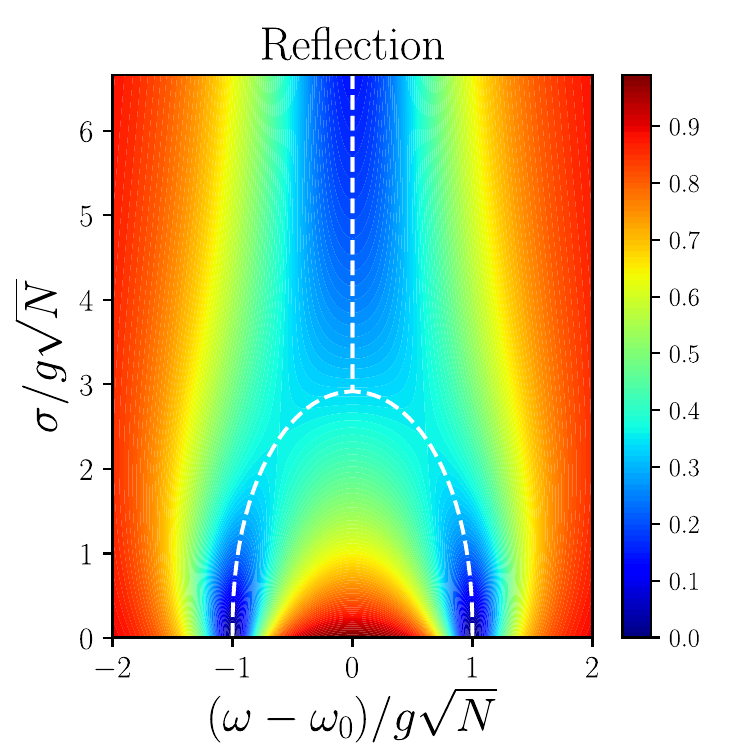}

}

\caption{Numerically calculated (A) absorption, (B) transmission, and (C) reflection
spectra for a Lorentzian distribution of excitation energies $p(\omega_{ex})$
centered at $\omega_{ex}=\omega_{0}$; $\omega_{ph}=\omega_{0}$ and
$\kappa_{L}=\kappa_{R}=\frac{1}{2}g\sqrt{N}$. The white dashed lines
indicate our analytical results for the polariton frequencies, showing
strong agreement with the calculated spectra overall magnitudes of
disorder. As done in Ref. \cite{sommer2021molecular}, the white solid
lines in (B) represent individual spectra for weak and strong disorder
systems, highlighting the transition from two distinct peaks to a
broad central peak as disorder increases. \label{fig:Lorentzian_spectra}}

\end{figure*}

For Lorentzian disorder, the real and imaginary parts of the susceptibility
are 
\begin{align}
\chi'(\omega) & =-g^{2}N\frac{(\omega-\omega_{0})}{(\omega-\omega_{0})^{2}+(\sigma/2)^{2}},\label{eq:chi1_Lorentzian}\\
\chi''(\omega) & =g^{2}N\frac{\sigma/2}{(\omega-\omega_{0})^{2}+(\sigma/2)^{2}}.\label{eq:chi2_Lorentzian}
\end{align}
From Eqs. \ref{eq:polariton_absorption} and \ref{eq:transmission},
the absorption, transmission, and reflection spectra are given by
\begin{align}
A(\omega) & =\frac{\kappa_{L}\sigma g^{2}N}{|(\omega-\omega_{ph}+i\frac{\kappa}{2})(\omega-\omega_{0}+i\frac{\sigma}{2})-g^{2}N|^{2}},\label{eq:absorption_lorentzian}\\
T(\omega) & =\frac{\kappa_{L}\kappa_{R}[(\omega-\omega_{0})^{2}+(\sigma/2)^{2}]}{|(\omega-\omega_{ph}+i\frac{\kappa}{2})(\omega-\omega_{0}+i\frac{\sigma}{2})-g^{2}N|^{2}},\label{eq:transmission_lorentzian}\\
R(\omega) & =1-\frac{\kappa_{L}\Big\{\kappa_{R}[(\omega-\omega_{0})^{2}+(\sigma/2)^{2}]+\sigma g^{2}N\Big\}}{|(\omega-\omega_{ph}+i\frac{\kappa}{2})(\omega-\omega_{0}+i\frac{\sigma}{2})-g^{2}N|^{2}}.\label{eq:reflection_lorentzian}
\end{align}
Figure \ref{fig:Lorentzian_spectra} presents the numerically calculated
spectra. 

To find the extrema of each spectra, we solve for $\frac{d}{d\omega}T(\omega)=0$,
$\frac{d}{d\omega}A(\omega)=0$, and $\frac{d}{d\omega}R(\omega)=0$
separately. Lorentzian disorder affords exact analytical expressions
for all regimes of $\sigma$. In the case where the photon mode is
resonant with the center of the Lorentzian distribution, \emph{i.e.},
$\omega_{ph}=\omega_{0}$, we find that the upper ($\omega_{+}^{A}$)
and lower ($\omega_{-}^{A}$) polariton peaks in the absorption are
located at the frequencies 
\begin{align}
\omega_{\pm}^{A} & =\omega_{0}\pm\mathfrak{I}\Big[\sqrt{\frac{1}{8}(\sigma^{2}+\kappa^{2})-g^{2}N}\Big].\label{eq:Lorentzian_up_lp_abs}
\end{align}
Similarly, the upper and lower polariton peaks in the transmission
spectrum are located at 
\begin{equation}
\omega_{\pm}^{T}=\omega_{0}\pm\mathfrak{I}\Big[\sqrt{\frac{\sigma^{2}}{4}-g^{2}N\sqrt{1+\frac{\sigma(\kappa+\sigma)}{2g^{2}N}}}\Big],\label{eq:Lorentzian_up_lp_trans}
\end{equation}
and the peaks in the reflection spectrum are at

\begin{align}
\omega_{\pm}^{R} & =\omega_{0}\pm\mathfrak{I}\Bigg[\Bigg(\frac{\sigma^{2}}{4}+\frac{\sigma g^{2}N}{\kappa_{R}} \nonumber \\
& -g^{2}N\sqrt{\frac{\sigma(\kappa+\sigma)(\sigma-\kappa+2\kappa_{R})}{4\kappa_{R}g^{2}N}+\Big(1+\frac{\sigma}{\kappa_{R}}\Big)^{2}}\Bigg)^{1/2}\Bigg].\label{eq:Lorentzian_up_lp_refl}
\end{align}
With prior information on the molecular disorder ($\sigma$) and cavity
linewidth ($\kappa$), one can use the above equations to extract
the correct value for the collective coupling $g\sqrt{N}$ from experimentally
obtained spectra. For the transmission spectra, we observe similar
behavior to that was shown in Refs. \cite{sommer2021molecular,gera2022effects,gera2022exact,cohn2022vibrational,zhou2023interplay}
for Gaussian disorder: the VRS initially increases with increasing
disorder; as disorder increases further, the VRS saturates, and then
decreases to zero. It is interesting that this trend is not observed
for absorption, as VRS only \emph{decreases} with disorder. Table
\ref{tab:Trends} summarizes the intricate behavior of VRS across
spectra for different disorder distributions. Qualitatively, the increase
of VRS with weak disorder is a manifestation of level repulsion (see
Figure \ref{fig:level_repulsion}). It should be noted that Eqs. \ref{eq:absorption_lorentzian}-\ref{eq:transmission_lorentzian}
are the same as those presented in Ref. \cite{savona1995quantum}. Furthermore,
the expressions for $\kappa=0$, Eq. \ref{eq:Lorentzian_up_lp_abs}
looks similar to the analytical expressions derived by Refs. \cite{engelhardt2022unusual,gera2022effects} and  \cite{climent2023kubo},
but the decrease in VRS for weak
disorder differs by a factor of a half compared to our result. This
is due to the fact that the real parts of the poles for Eqs. \ref{eq:Lorentzian_up_lp_abs}-\ref{eq:Lorentzian_up_lp_refl}
do not correspond to the true extrema along the real-value frequency
axis.

\begin{table*}
\centering
\caption{Vacuum Rabi splitting trends with disorder \label{tab:Trends}}
\begin{tabular}{ccccccccc}
 
\multicolumn{1}{l}{} & \multicolumn{2}{c}{{\footnotesize{}Gaussian}} &  & \multicolumn{2}{c}{{\footnotesize{}Lorentzian}\textsuperscript{}} &  & \multicolumn{2}{c}{{\footnotesize{}Rectangle}}\\
\cline{2-9}
\noalign{\vskip0.1cm}
{\footnotesize{}Optical Signal} & {\footnotesize{}$\text{\ensuremath{\sigma<g\sqrt{N}}}$} & {\footnotesize{}$\sigma>g\sqrt{N}$} &  & {\footnotesize{}$\text{\ensuremath{\sigma<g\sqrt{N}}}$} & {\footnotesize{}$\sigma>g\sqrt{N}$} &  & {\footnotesize{}$\text{\ensuremath{\sigma<g\sqrt{N}}}$} & {\footnotesize{}$\sigma>g\sqrt{N}$}\\
\midrule
{\footnotesize{}$A(\omega)$} & {\footnotesize{}increases} & {\footnotesize{}decreases} &  & {\footnotesize{}decreases} & {\footnotesize{}decreases} &  & {\footnotesize{}increases} & {\footnotesize{}decreases; narrow side bands}\\
{\footnotesize{}$T(\omega)$} & {\footnotesize{}increases} & {\footnotesize{}decreases} &  & {\footnotesize{}increases} & {\footnotesize{}decreases} &  & {\footnotesize{}increases} & {\footnotesize{}decreases; narrow side bands}\\
{\footnotesize{}$R(\omega)$} & {\footnotesize{}increases} & {\footnotesize{}decreases} &  & {\footnotesize{}increases or decreases}\textsuperscript{{\footnotesize{}a}} & {\footnotesize{}decreases} &  & {\footnotesize{}increases} & {\footnotesize{}decreases; narrow side bands}\\
\midrule
\multicolumn{9}{l}{{\footnotesize{}}\textsubscript{}{\footnotesize{}}\textsuperscript{{\footnotesize{}a}}{\footnotesize{}For
the Lorenzian, the VRS in the reflection spectrum increases if $\kappa_{L}^{2}/\kappa_{R}^{2}<1$
and decreases if $\kappa_{L}^{2}/\kappa_{R}^{2}>1$ (See Table \ref{tab:Vacuum-Rabi-Splitting})}}\\
\end{tabular}
\end{table*}

\subsection{Gaussian and rectangular disorder}
For Gaussian and Rectangular disorder, we can still extract semi-analytical
results when $\sigma\ll g\sqrt{N}$ (weak disorder) and $|\omega-\omega_{0}|\approx g\sqrt{N}$
(about the polariton peaks). For the Gaussian distribution, we employ
the asymptotic expansion of the Dawson function \cite{hummer1964expansion},
similar to Ref. \cite{gera2022effects}, up to $O[(\frac{\sigma}{\omega-\omega_{0}})^{3}]$
to obtain an approximate expression for the real part of the susceptibility,
\begin{align}
\chi'(\omega) & \approx-g^{2}N\Bigg[\frac{1}{\omega-\omega_{0}}+\frac{\sigma^{2}}{(\omega-\omega_{0})^{3}}\Bigg].\label{eq:asymptotic_expansion}
\end{align}
Meanwhile, for the rectangular distribution, we get,
\begin{equation}
\chi'(\omega)\approx-g^{2}N\Bigg[\frac{1}{\omega-\omega_{0}}+\frac{\sigma^{2}}{12(\omega-\omega_{0})^{3}}\Bigg].\label{eq:taylor_expansion_rectangle}
\end{equation}
Note that for both distributions $\chi''(\omega)\approx0$ at the
polaritonic windows, so no VRS is predicted in the absorption spectrum
for both Gaussian and rectangular distributions. The lack of VRS in
the absorption spectrum for these disorder distributions is due to
the minimal overlap between the molecular absorption spectrum and
the polariton transmission, since the tails of the Gaussian die quickly
away from $\omega_{0}$, while the rectangular distribution has no
tails. Contrast this observation with the analogous one for the Lorentzian
distribution, which does present VRS in its absorption spectrum owing
to the long tails of the Lorentzian.

\begin{figure*}[!ht]
\subfloat[]{\includegraphics[width=0.666\columnwidth]{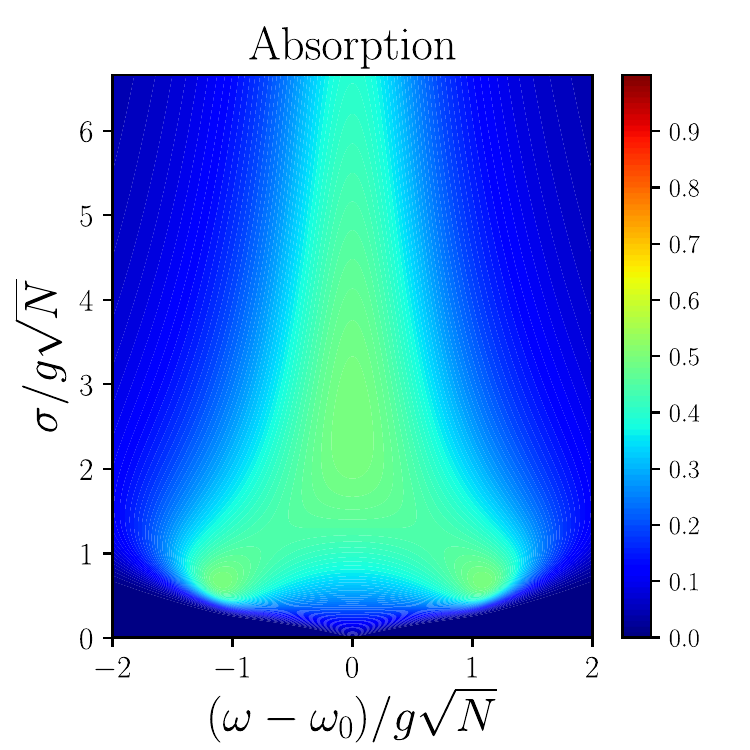}

}\subfloat[]{\includegraphics[width=0.666\columnwidth]{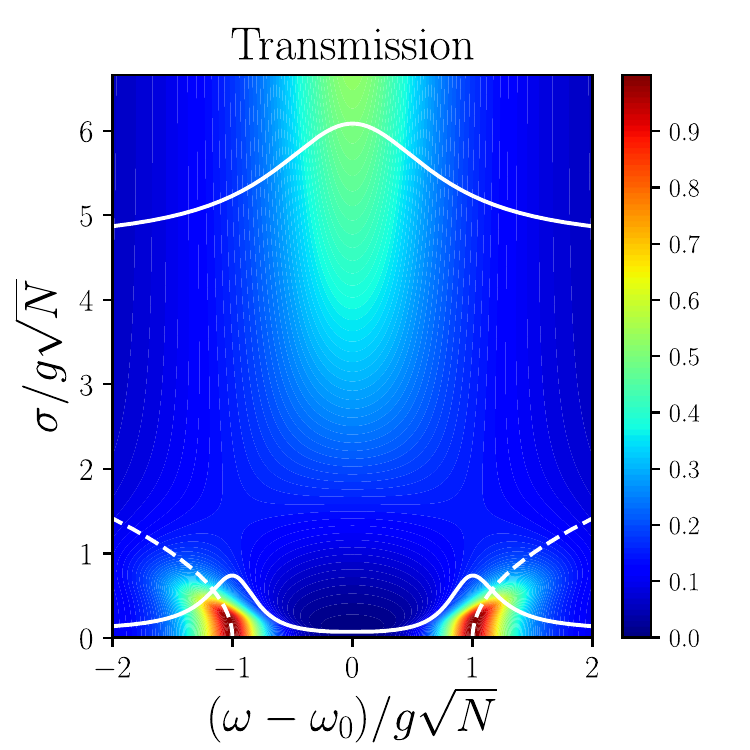}

}\subfloat[]{\includegraphics[width=0.666\columnwidth]{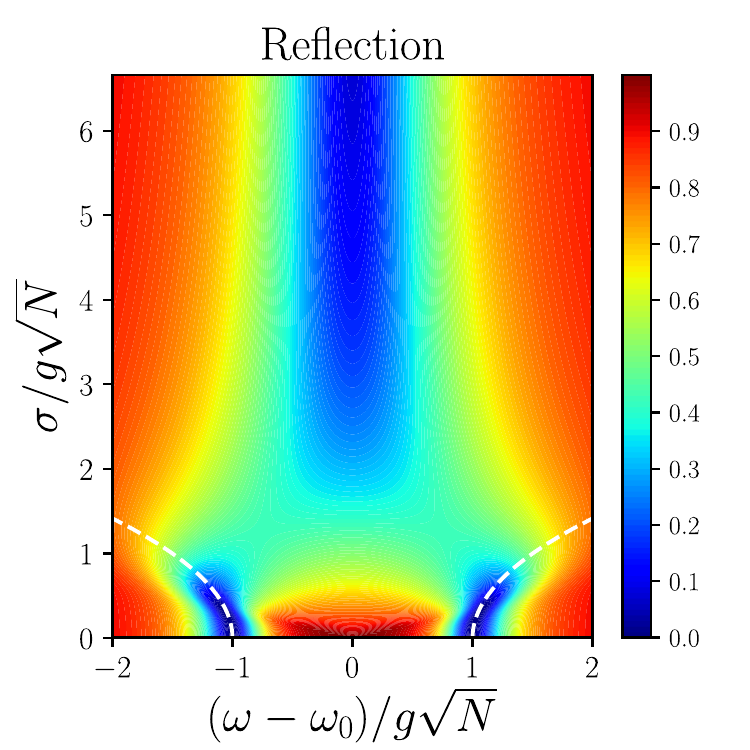}

}

\caption{Numerically calculated (A) absorption, (B) transmission, and (C) reflection
spectra for a Gaussian distribution of excitation energies $p(\omega_{ex})$
centered at $\omega_{ex}=\omega_{0}$; $\omega_{ph}=\omega_{0}$ and
$\kappa_{L}=\kappa_{R}=\frac{1}{2}g\sqrt{N}$. The white dashed lines
indicate our analytical results for the polariton frequencies, showing
strong agreement with the calculated spectra for weak disorder. As
done in Ref. \cite{sommer2021molecular}, the white solid lines in
(B) represent individual spectra for weak and strong disorder systems,
highlighting the transition from two distinct peaks to a broad central
peak as disorder increases.\label{fig:Gaussian_spectra}}
\end{figure*}

Using these approximations, we find that for $\omega_{ph}=\omega_{0}$,
the transmission and reflection spectra for Gaussian distribution
are approximately 
\begin{align}
T(\omega) & \approx\frac{\kappa_{L}\kappa_{R}g^{6}N^{3}}{|(\omega-\omega_{0})^{4}-g^{2}N(\omega-\omega_{0})^{2}-g^{2}N\sigma^{2}+ig^{3}N^{\frac{3}{2}}\frac{\kappa}{2}|^{2}},\label{eq:approx_gaussian_transmission}\\
R(\omega) & \approx1-T(\omega),\label{eq:approximate_gaussian_refleciton}
\end{align}
with the polariton frequencies for both spectra approximately located
at
\begin{equation}
\omega_{\pm}=\omega_{0}\pm\sqrt{\frac{1}{2}g^{2}N+\frac{1}{2}\sqrt{g^{4}N^{2}+4g^{2}N\sigma^{2}}}.\label{eq:gaussian_UP_LP}
\end{equation}
Similarly, we find that the transmission and reflection spectra for
the rectangular distribution are approximately 
\begin{align}
T(\omega) & \approx\frac{\kappa_{L}\kappa_{R}g^{6}N^{3}}{|(\omega-\omega_{0})^{3}-g^{2}N(\omega-\omega_{0})-\frac{g^{2}N\sigma^{2}}{12}+ig^{3}N^{\frac{3}{2}}\frac{\kappa}{2}|^{2}},\label{eq:Transmission_rect}\\
R(\omega) & \approx1-T(\omega).\label{eq:reflection_rect}
\end{align}
In this case the polariton frequencies are approximately located at
\begin{equation}
\omega_{\pm}=\omega_{0}\pm\sqrt{\frac{1}{2}g^{2}N+\frac{1}{2}\sqrt{g^{4}N^{2}+\frac{\sigma^{2}g^{2}N}{3}}},\label{eq:rectangle_up_lp}
\end{equation}
which is similar to Eq. \ref{eq:gaussian_UP_LP}.

Figure \ref{fig:Gaussian_spectra} shows the numerically calculated
spectra for Gaussian disorder. The trend in VRS in the transmission
and reflection spectra as a function of $\sigma$ is qualitatively
the same as that for the Lorentzian. Eq. \ref{eq:gaussian_UP_LP}
is in good agreement with the numerical results for $\sigma\ll g\sqrt{N}$,
and upon Taylor expanding around $\sigma=0$ to second order, reduces
to the analytical results of Refs. \cite{gera2022effects} and \cite{zhou2023interplay}.
The analytical results of Refs. \cite{sommer2021molecular} and \cite{cohn2022vibrational}
qualitatively capture the behavior of the transmission spectrum; however,
quantitatively they fits better with the Lorentzian disorder. The
transformation of Gaussian into Lorentzian disorder in these references
appears to be due to the usage of the Markovian approximation. Figure
\ref{fig:Gaussian_spectra} also highlights the dangers for taking
VRS at face value. We observe that the largest VRS in the transmission
and reflection spectra is approximately $1.5\times2g\sqrt{N}$. This
implies that one must be careful using VRS to determine the strength
of the collective light-matter coupling. For example, if $\text{VRS}/2\omega_{0}\approx0.1$
one may mistakenly claim to be in the ultrastrong coupling regime,
while in reality the collective light-matter coupling $g\sqrt{N}\approx0.07$
is still within the strong-coupling regime. 
\begin{figure*}[!ht]
\subfloat[]{\includegraphics[width=0.666\columnwidth]{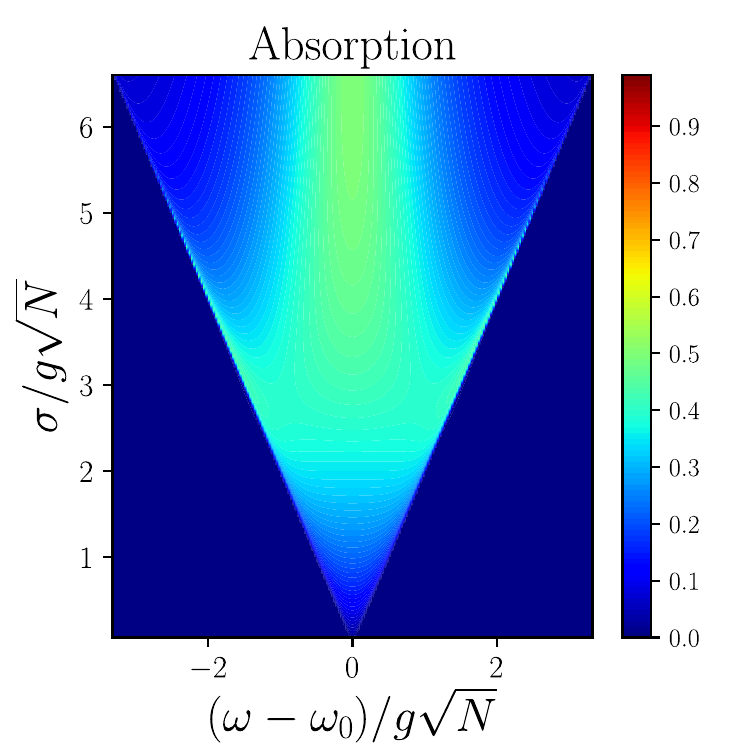}

}\subfloat[]{\includegraphics[width=0.666\columnwidth]{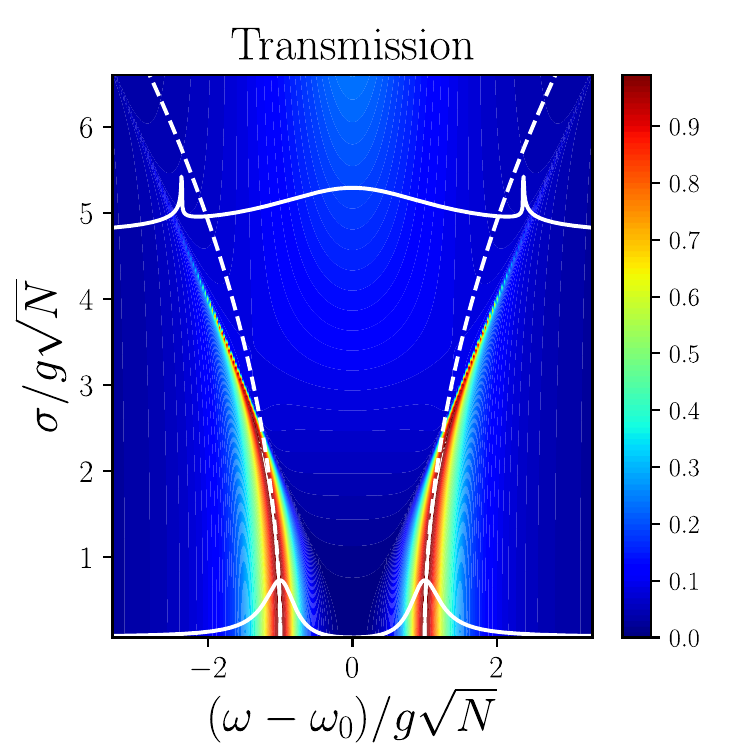}

}\subfloat[]{\includegraphics[width=0.666\columnwidth]{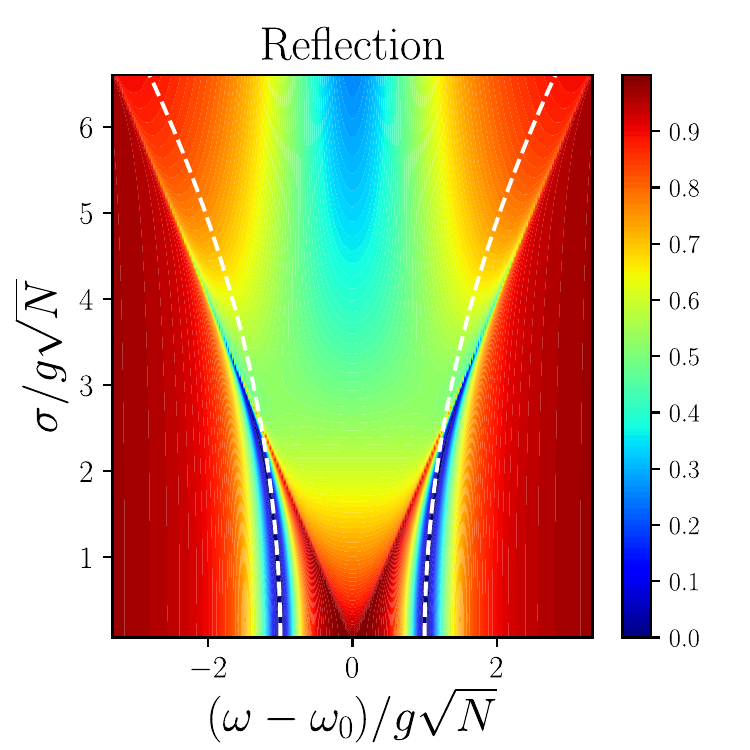}

}

\caption{Numerically calculated (A) absorption, (B) transmission, and (C) reflection
spectra for a rectangular distribution of excitation energies $p(\omega_{ex})$
centered at $\omega_{ex}=\omega_{0}$; $\omega_{ph}=\omega_{0}$ and
$\kappa_{L}=\kappa_{R}=\frac{1}{2}g\sqrt{N}$. The white dashed lines
indicate our analytical results for the polariton frequencies, showing
strong agreement with the calculated spectra for weak disorder. As
done in Ref. \cite{sommer2021molecular}, the white solid lines in
(B) represent individual spectra for weak and strong disorder systems,
highlighting the transition from two distinct peaks to a broad central
peak as disorder increases. Remarkably, for large disorder, two peaks
that are narrower than the line width of the cavity and molecular
disorder emerge on either side of the broad central peak. \label{fig:Rectangular_spectra}}
\end{figure*}

Figure \ref{fig:Rectangular_spectra} displays the numerically calculated
spectra for the rectangular disorder. Such a scenario was briefly
considered in Ref \cite{gera2022effects}; here, we provide further
analysis. At low disorder, the rectangular distribution exhibits similar
behavior to the Gaussian distribution, as predicted by Eq. \ref{eq:rectangle_up_lp}.
As disorder increases, a broad central peak forms, consistent with
both the Lorentzian and Gaussian distributions. Intriguingly, as disorder
increases, two sharp sidebands for the polaritons also emerge, each
narrower than the cavity linewidth and the width of the rectangular
distribution, respectively. This is reminiscent to the pronounced
spectral narrowing witnessed in plasmonic surface lattice resonances
\cite{kravets2018plasmonic}. The unique characteristics of the rectangular
disorder, including singularities in its real part of the susceptibility
$\chi'(\omega)$ near these polariton peaks (see Table \ref{tab:Real-and-imaginary}),
are responsible for this phenomenon. The reduced linewidth of these
peaks indicates a higher degree of coherence lifetime within the system,
even in the presence of significant energetic disorder. Consequently,
this observation presents a promising avenue for applications in molecular
polariton systems requiring prolonged coherences between cavity and
molecules. In a realistic experiment, this phenomenon will only occur
if the underlying members of the disorder distribution have a small
Lorentzian homogeneous linewidth (which has not been treated in this
work); a large such linewidth will smoothen the singularity and reduce
the problem to the Lorentzian case.

\subsection{Sum rule}
From our extensive study of disorder effects on the polariton linear
spectrum, it is evident from the three disorder distributions studied
in this work, that disorder significantly impacts the value of VRS,
seemingly posing a challenge in extracting the precise value of the
collective light-matter coupling. However, a simple sum rule can be
utilized when both the absorption and transmission polariton spectra
are accessible. By integrating the ratio of these two signals ($I$),
which is proportional to the imaginary part of the linear molecular
susceptibility $\chi''(\omega)=g^{2}N\pi p(\omega)$, a robust method
for determining the collective light-matter coupling is revealed (see
Eqs. \ref{eq:polariton_absorption} and \ref{eq:transmission}): 
\begin{equation}
I=\int d\omega\frac{A(\omega)}{T(\omega)}=\frac{2\pi}{\kappa_{R}}g^{2}N.\label{eq:sum_rule}
\end{equation}
Importantly, this expression is general for any form and strength
of disorder. Note that Eq. \ref{eq:sum_rule} is presented in the
low temperature limit. For finite temperature effects, see Appendix. Therefore, experimental setups that enable the measurement
of both absorption and transmission spectra are deemed ideal for extracting
accurate collective-light-matter coupling values.

\begin{table*}[!ht]
\centering
\caption{Vacuum Rabi splitting expressions for when $\sigma\ll g\sqrt{N}$\label{tab:Vacuum-Rabi-Splitting}}
\begin{tabular}{cccc}

 & \multicolumn{3}{c}{Vacuum Rabi Splitting}\tabularnewline
\cline{2-4}
Optical Signal & Gaussian & Lorentzian\textsuperscript{a} & Rectangle\\
\midrule 
\noalign{\vskip0.5cm}
$A(\omega)$ & 0 & $2g\sqrt{N}-\frac{\sigma^{2}}{8\sqrt{g^{2}N-\frac{\kappa^{2}}{8}}}$ & 0\\
\noalign{\vskip0.5cm}
$T(\omega)$ & $2\sqrt{\frac{1}{2}g^{2}N+\frac{1}{2}\sqrt{g^{4}N^{2}+4g^{2}N\sigma^{2}}}$ & $2g\sqrt{N}+\frac{\kappa\sigma}{4g\sqrt{N}}$ & $2\sqrt{\frac{1}{2}g^{2}N+\frac{1}{2}\sqrt{g^{4}N^{2}+\frac{\sigma^{2}g^{2}N}{3}}}$\\
\noalign{\vskip0.5cm}
$R(\omega)$ & $2\sqrt{\frac{1}{2}g^{2}N+\frac{1}{2}\sqrt{g^{4}N^{2}+4g^{2}N\sigma^{2}}}$ & $2g\sqrt{N}-\frac{(\frac{\kappa_{L}^{2}}{\kappa_{R}}-\kappa_{R})\sigma}{8g\sqrt{N}}$ & $2\sqrt{\frac{1}{2}g^{2}N+\frac{1}{2}\sqrt{g^{4}N^{2}+\frac{\sigma^{2}g^{2}N}{3}}}$\\
\noalign{\vskip0.5cm}
\midrule
\noalign{\vskip0.1cm}
\multicolumn{4}{c}{\textsuperscript{a}For larger values of $\sigma$, Eqs. \ref{eq:Lorentzian_up_lp_abs}-\ref{eq:Lorentzian_up_lp_refl}
should be used to extract the correct value for $g\sqrt{N}$.}\\
\end{tabular}
\end{table*}

\section{Conclusions}

Our comprehensive investigation has unraveled the intricate relationship
between molecular disorder and VRS in molecular polariton. We have
derived precise analytical expressions for absorption, transmission,
and reflection spectra across various disorder distributions. Furthermore,
our study introduces a generalized sum rule for determining the collective
light-matter coupling under any form of disorder. These findings do
not only clarify the nuanced behavior of VRS amidst disorder but also
establish a reliable framework for extracting light-matter coupling
parameters with high accuracy from experimental data. In practical
terms, when the experimental set-up allows access to both transmission
and absorption signals, the sum rule can be readily applied. In situations
where accessing both signals is not possible, researchers can leverage
Eq. \ref{eq:polariton_absorption}-\ref{eq:refleciton}, coupled with
the analytical forms of the molecular susceptibility outlined in Table
\ref{tab:Real-and-imaginary}, to effectively fit their experimental
results. Both these approaches ensure the extraction of the correct
value of $g\sqrt{N}$ in the presence of disorder of any magnitude.
Additionally, for scenarios involving mild disorder, the simplified
expressions provided in Table \ref{tab:Vacuum-Rabi-Splitting} offer
a convenient solution for data fitting. Furthermore, our study has
unveiled a fascinating phenomenon associated with rectangular disorder
-- the emergence of narrow sidebands alongside a broad central peak.
This intriguing line narrowing, observed in the presence of significant
disorder, suggests a higher degree of coherence lifetime within the
system. Such behavior is especially promising for applications requiring
long-lived coherences between the cavity and molecules, providing
an exciting avenue for future research in the realm of molecular polaritons.

\begin{acknowledgement}
The authors thank St{\'e}phane K{\'e}na-Cohen for suggesting that a sum rule should be the most precise way to extract accurate VRSs. Additionally, the authors acknowledge the use of ChatGPT for assistance with sentence editing. The text was then proofread by the authors and revised if necessary.  K. S. thanks Arghadip Koner and Michael Reitz for useful discussions. 
\end{acknowledgement}

\begin{funding}
This work was supported by the Air Force Office of Scientific Research (AFOSR) through the Multi-University Research Initiative (MURI) program no. FA9550-22-1-0317. 
\end{funding}

\begin{authorcontributions}
K. S. developed the model and calculations,
N. C. G. and J. Y.-Z. guided the conceptualization of
the disorder profiles and sum rule, and J.Y.-Z. supervised
the work throughout. All authors have accepted responsibility
for the entire content of this manuscript and approved its
submission.
\end{authorcontributions}

\begin{conflictofinterest}
Authors state no conflict of interest.
\end{conflictofinterest}

\begin{dataavailabilitystatement}
Data underlying the results presented
in this paper may be obtained from the corresponding
author upon reasonable request.

\end{dataavailabilitystatement}

\appendix
\renewcommand{\theequation}{A\arabic{equation}}
\setcounter{equation}{0}
\section{Sum rule for finite temperature}
From Eqs. \ref{eq:polariton_absorption} and \ref{eq:transmission},
we have that 
\begin{equation}
I=\int d\omega\frac{A(\omega)}{T(\omega)}=\frac{2}{\kappa_{R}}\int d\omega\chi''(\omega)\label{eq:sum_rule_integral_SI}
\end{equation}
From Eq. \ref{eq:molecular_suceptabilty}, for $N$ two-level systems
coupled to a single cavity mode at a given temperature $T$, Eq. \ref{eq:sum_rule_integral_SI}
becomes:
\begin{align}
I&=\frac{i}{\kappa_{R}}\lim_{\gamma\to0^{+}}\sum_{i=1}^{N}\tanh\Big(\frac{\hbar\omega_{ex,i}}{2k_{B}T}\Big)|\lambda\mu_{i}|^{2} \nonumber \\
&\times \int d\omega\Big[\frac{1}{\omega-\omega_{ex,i}+i\frac{\gamma}{2}}-\frac{1}{\omega-\omega_{ex,i}-i\frac{\gamma}{2}}\Big]\label{eq:sum_rule_integral_arb_T_SI}
\end{align}
 Note that if we evaluate the integral counter clockwise over a semicircle
$S_{+}$ with an infinite radius in the upper half of the complex
plane, we have that \begin{subequations}
\begin{align}
0 & =\oint_{S_{+}}d\omega\Big[\frac{1}{\omega-\omega_{ex,i}+i\frac{\gamma}{2}}] \nonumber \\
&=\int d\omega\Big[\frac{1}{\omega-\omega_{ex,i}+i\frac{\gamma}{2}}\Big]+i\pi\label{eq:contour_integral_1}\\
i2\pi & =\oint_{S_{+}}d\omega\Big[\frac{1}{\omega-\omega_{ex,i}-i\frac{\gamma}{2}}] \nonumber \\
&=\int d\omega\Big[\frac{1}{\omega-\omega_{ex,i}-i\frac{\gamma}{2}}\Big]+i\pi\label{eq:contour_integral_2}
\end{align}
\end{subequations} This implies that $\int d\omega\Big[\frac{1}{\omega-\omega_{ex,i}+i\frac{\gamma}{2}}\Big]=-i\pi$
and $\int d\omega\Big[\frac{1}{\omega-\omega_{ex,i}-i\frac{\gamma}{2}}\Big]=i\pi$.
Substituting these values into Eq. \ref{eq:sum_rule_integral_arb_T_SI},
we arrive at the generalized sum rule for arbitrary $T$:
\begin{equation}
I=\frac{2\pi}{\kappa_{R}}\sum_{i=1}^{N}\tanh\Big(\frac{\hbar\omega_{ex,i}}{2k_{B}T}\Big)|\lambda\mu_{i}|^{2}=\frac{2\pi}{\kappa_{R}}Ng_{eff}^{2}\label{eq:sum_rule_arbritrary_N}
\end{equation}
 where 
\begin{equation}
g_{eff}^{2}=\int d\omega_{ex}p(\omega_{ex})\tanh\Big(\frac{\hbar\omega_{ex}}{2k_{B}T}\Big)|\lambda\mu(\omega_{ex})|^{2}\label{eq:g_eff_squared}
\end{equation}
is the square of the effective single molecule light-matter coupling.
Here $p(\omega_{ex})$ is the probability distribution of excitation
frequencies. Note in the limit that $\hbar\omega_{ex,i}\gg k_{B}T$
and $\mu(\omega_{ex})\to\mu$, we have that $g_{eff}^{2}\to g^{2}$,
recovering the version of the sum rule presented in the main text.



\end{document}